\documentclass[preprint, 3p, twocolumn]{elsarticle}
\usepackage{graphicx}
\usepackage{booktabs}
\usepackage{tabularx}
\usepackage[normalem]{ulem}
 \useunder{\uline}{\ul}{}
 \usepackage{rotating}
 \usepackage{dsfont}
 \usepackage{textcomp}
 \usepackage{upgreek}

\usepackage{color}
\def\Manu#1{\textcolor{black}{#1}}
\usepackage{tabularx}
\newcolumntype{Y}{>{\centering\arraybackslash}X}
\usepackage[switch]{lineno}
\usepackage{graphicx,xcolor}
\usepackage{epstopdf}
\usepackage{setspace}
\usepackage[T1]{fontenc}
\usepackage{amssymb,amsfonts,amsmath}
\usepackage{setspace}
\usepackage{subcaption}
\usepackage{mathtools}
\usepackage{etoolbox}
\biboptions{sort&compress}

\usepackage{hyperref}
\hypersetup{
    colorlinks=true,
    linkcolor=blue,
    filecolor=magenta,      
    urlcolor=cyan,
}
\begin{document}

\title{\Manu{Biomolecular condensates and their pathologies: A colloidal, interfacial, and transport perspective}}
\author[vgodavarti]{V. Godavarti}
\author[vinny]{V. Chandran Suja}
\address[vgodavarti]{Westford Academy, 30 Patten Road, Westford, MA, 01886}
\address[vinny]{School of Engineering and Applied Sciences, Harvard University, Boston, Massachusetts, USA}

\begin{abstract}
Biomolecular condensates are micron-scale, membraneless compartments that concentrate proteins and nucleic acids by liquid-liquid phase separation (LLPS). Condensates play critical roles in transcription, mRNA metabolism, stress responses, and signaling across the cell, with their dysregulation increasingly correlated with several pathologies spanning neurodegeneration, cancer, cardiometabolic disease, and viral infections. The emergent biology of condensates can be cohesively understood through the framework of colloidal, interfacial, and transport phenomena. Here we review the condensate literature through this joint lens, treating each condensate as a biological colloid whose behavior is captured by four measurable physical quantities: a Flory--Huggins interaction parameter that determines whether the dense phase forms at all, a partition coefficient that dictates the degree of client molecule accumulation, a viscoelastic relaxation time that sets whether the dense phase responds as a liquid or a solid, and an interfacial tension that governs droplet shape and interfacial dynamics. Each quantity is traced to its physical origin, its measured range in biological condensates, and the techniques that quantify it, for both three-dimensional droplets and two-dimensional membrane-tethered condensates. Interpreting disease as the displacement of these physical quantities from their functional ranges, we reorganize condensate pathologies into a small set of physical failure modes --- saturation concentration dysregulation, aberrant aging and liquid-to-solid transitions, disrupted partitioning, interfacial aggregation, and pathogen hijacking --- and show how the same parameters become design variables for condensate-targeted drugs and engineered coacervate delivery vehicles. The review concludes by highlighting open questions at the interface of condensate biology and colloidal and interfacial science.
\end{abstract}

\begin{keyword}
  Biomolecular condensates \sep liquid-liquid phase separation \sep complex colloids \sep complex coacervation \sep macromolecular crowding \sep intracellular transport \sep membraneless organelles
\end{keyword}

\maketitle

\section{Introduction}\label{sec:Introduction}
Biomolecular condensates are micron-scale structures that concentrate biomaterials inside cells without enclosing them in a lipid membrane \cite{banani2017biomolecular,hyman2014liquid}. Unlike classical membrane-bound organelles, condensates form via reversible liquid-liquid phase separation of biological polymers like proteins and nucleic acids, and are distributed throughout the cell in the form of three- and two-dimensional (membrane-tethered) structures. They are vital to cellular functioning, playing important roles in processes including transcription, splicing, metabolism, signaling, and stress responses. Condensate dysregulation is also increasingly attributed to pathologies including neurodegeneration, cancer, and cardiometabolic diseases, thus presenting a novel therapeutic target for these challenging diseases. Despite their biological complexity, condensates are now understood to be colloidal objects whose stability, internal structure, and dynamics are governed by the colloidal, interfacial, and transport principles that drive coexistence in multicomponent polymeric solutions. 

Two separate lines of research converged on this view over the last century. The first originated from efforts to understand the spontaneous separation of homogeneous polymeric solutions into coexisting liquid phases with distinct composition. This process, widely referred to as liquid--liquid phase separation (LLPS), was first reported by Beijerinck in 1896, who showed that mixtures of gelatin and starch separate into two liquid phases, each enriched in one of the polymers \cite{beijerinck1896}. The charge-mediated subset of LLPS, in which oppositely charged biopolymers associate to form a dense, polymer-rich phase, was investigated in the 1920s and 1930s by Bungenberg de Jong and Kruyt, who named the process \emph{complex coacervation} \cite{bungenberg1929}. The thermodynamic basis for LLPS in polymer solutions was placed on a quantitative footing in 1942 by Flory and Huggins, whose independent lattice mean-field treatments cast the competition between mixing entropy and a single polymer--solvent interaction parameter $\chi$ as a coexistence curve sensitive to polymer chain length and temperature \cite{flory1942thermodynamics,huggins1942theory}. Successive refinements over the following half century extended this framework along directions relevant to biological polymers: Overbeek and Voorn introduced the electrostatic correlations needed to describe complex coacervation of oppositely charged polyelectrolytes \cite{overbeek1957phase}; de Gennes formalized the scaling behavior of flexible chains in good and poor solvents and their conformational crossovers in semidilute and concentrated regimes \cite{degennes1979scaling}; Doi and Edwards systematized the dynamical and viscoelastic response of entangled polymer melts and networks \cite{doi1986theory}; Cahn placed critical-point wetting at fluid and solid interfaces on a thermodynamic basis \cite{cahn1977critical}; while Semenov and Rubinstein formalized the sticker-and-spacer paradigm to capture the multivalent reversible interactions in associative polymers \cite{semenov1998thermoreversible}. On the experimental side, a rich colloidal science toolkit for measuring interfacial tension, partition coefficients, viscoelastic moduli, and wetting angles was developed and refined into a form suitable for studying biopolymeric solutions \cite{ryden1971interfacial,albertsson1971partition,mason1995optical,degennes1985wetting, adamson1997physical}. In essence, by the end of the twentieth century, the physical-chemistry vocabulary and tools needed to describe the condensation of biological polymers were largely in place.

The biological recognition of membraneless cellular structures developed along an entirely separate line. The nucleolus was first described as a discrete intranuclear body in the 1830s \cite{mitrea2022modulating}. By 1903, both Cajal bodies and viral Negri bodies had also been described as distinct intracellular structures \cite{gall2003centennial,nevers2020negri}, although their membraneless nature — and their functions — would only be established decades later with the advent of electron microscopy. Across the twentieth century, the catalog of such membraneless organelles expanded steadily to include P-bodies, stress granules, germ granules, signaling puncta, and an array of nuclear bodies, all sharing the property of concentrating specific proteins and nucleic acids without an enclosing bilayer. Functional studies linked these compartments to specific cellular roles: ribosome biogenesis in the nucleolus, mRNA storage and turnover in P-bodies, and sequestration of stalled initiation complexes in stress granules \cite{boisvert2007multifunctional,parker2007p,buchan2009eukaryotic}. Concurrent research also established the dynamic nature of membraneless organelles, characterizing how they assemble and disassemble in response to nutrient availability, oxidative stress, viral infection, and cell-cycle progression. What remained absent from the biological account, until recently, was a physical mechanism that could explain how these cellular structures maintained a defined boundary, a stable composition, and a finite lifetime without an enclosing bilayer. 

The two lines of research converged in 2009, when a seminal study by Brangwynne and colleagues showed that P granules of \textit{C.~elegans} embryos behave as liquid droplets with a measurable viscosity and interfacial tension \cite{brangwynne2009germline}. In doing so, this study established that membraneless organelles are in fact phase-separated entities whose formation, structure, and dynamics are described by the polymer-physics toolkit developed over the preceding century. The subsequent decade saw an explosion of work mapping the chemistry, dynamics, and function of biomolecular condensates across cell biology \cite{li2012phase,hyman2014liquid,banani2017biomolecular,shin2017liquid,alberti2021biomolecular,boeynaems2018protein}, with extensive measurements of biophysical properties of condensates including their interfacial tension \cite{jawerth2018salt}, viscoelastic relaxation \cite{jawerth2020protein}, and partition coefficients \cite{klein2020partitioning}. The remainder of this review develops that connection systematically. Section~\ref{sec:Types} surveys the structural taxonomy of three-dimensional cytoplasmic and nucleoplasmic condensates and the two-dimensional membrane-tethered clusters. Section~\ref{sec:Colloidal} treats the colloidal and interfacial physics that governs condensate formation, structure, and stability---Flory--Huggins coexistence, sticker-and-spacer chemistry, interfacial tension, electrostatic boundary effects, macromolecular crowding, viscoelastic aging, and membrane wetting. Section~\ref{sec:Transport} treats the transport phenomena that move material into, through, and between condensates, including partitioning, dense-phase diffusion, interfacial mass transfer, Marangoni and diffusiophoretic forcing, and active driving. With this physical machinery in place, Section~\ref{sec:Pathologies} reorganizes the condensate-pathology literature into a small set of physical failure modes, each corresponding to an identifiable physical parameter excursion. Section~\ref{sec:Therapeutic} uses the same parameters as design variables for both condensate-targeted therapeutics and engineered coacervate delivery platforms. The review concludes by highlighting outstanding questions at the interface of condensate biology and colloidal and interfacial science.

\begin{figure*}[!ht]
\vspace{-0.75 cm}
    \centering
    \includegraphics[width=0.9\linewidth]{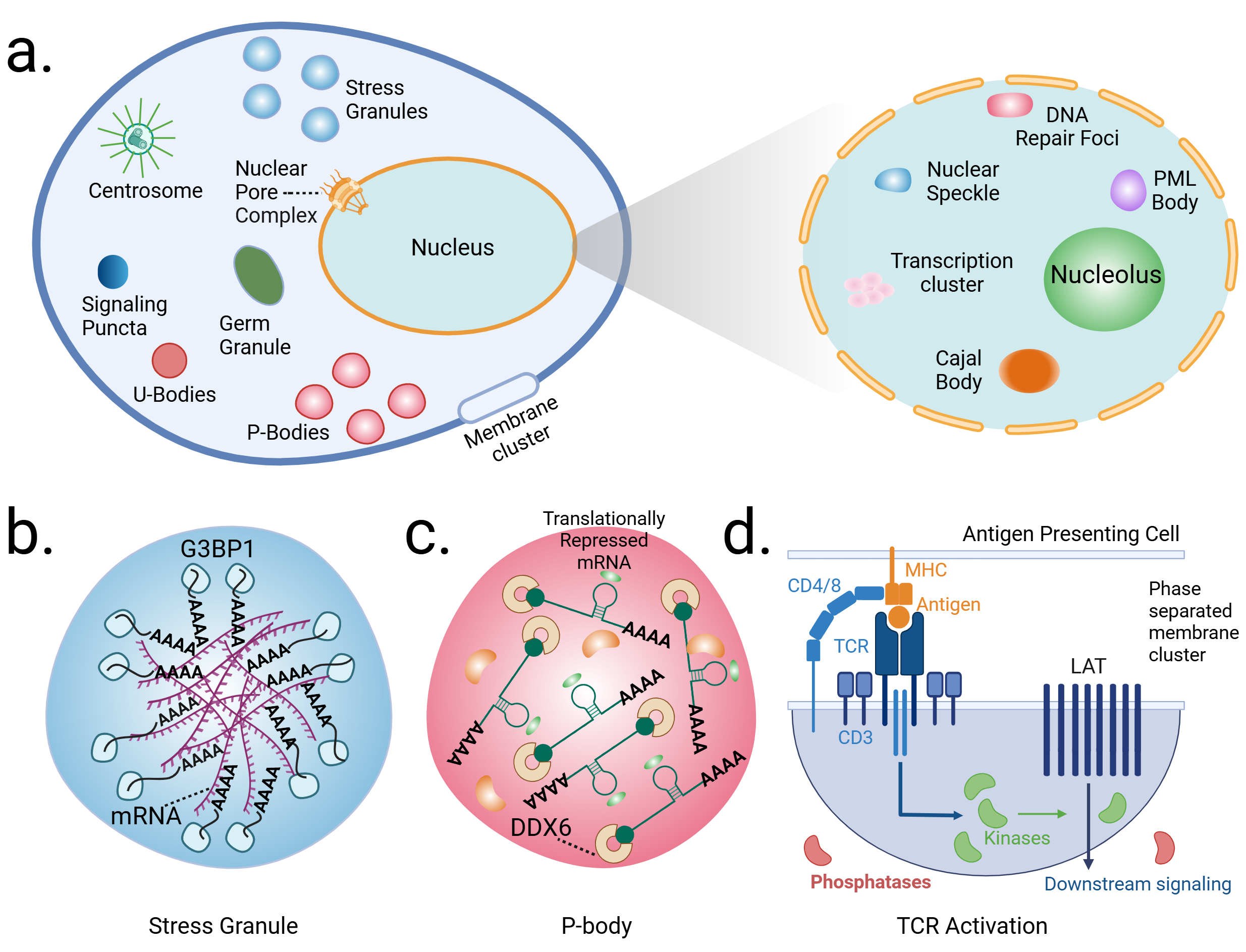}
    \caption{Biomolecular condensates in the cell. {\bf a.} A multitude of three-dimensional condensates populate the cytoplasm (stress granules, P-bodies, germ granules, U-bodies, and signaling puncta) and the nucleus (nucleolus, Cajal bodies, PML bodies, nuclear speckles, transcription clusters, and DNA repair foci). Two-dimensional condensates, phase separated clusters tethered to the plasma membrane, are also vital to cellular functioning. {\bf b.} Schematic of a stress granule that depicts RNA-binding protein G3BP1 cross-linking mRNA into a multivalent network. {\bf c.} A schematic of a P-body showing the DEAD-box helicase DDX6 binding translationally repressed mRNAs and, with its partner proteins, assembling them into a multivalent network. {\bf d.} A T-cell receptor (TCR) signaling cluster. Antigen presentation drives LAT to phase separate on the inner leaflet into a two-dimensional cluster that concentrates kinases and excludes phosphatases, sustaining downstream signaling.}
    \label{fig:intro}
\end{figure*}

\section{Types of biomolecular condensates}\label{sec:Types}
Diverse families of three- and two-dimensional (membrane-tethered) condensates reside in various parts of the cell and serve a variety of critical cellular functions (Fig.\ref{fig:intro}a). 

\subsection{Three-dimensional condensates}\label{sec:Types3D}
Stress granules form in response to various forms of cellular stress, such as extreme temperature fluctuations, heat shock, oxidative stress and virus infections. They are mainly composed of multifunctional RNA-binding proteins such as G3BP1 and translation initiation complexes that have stalled (Fig.\ref{fig:intro}b), enabling the cell to sequester and preserve mRNA until conditions improve \cite{protter2016principles,buchan2009eukaryotic}. Phosphorylation of eukaryotic initiation factor 2$\alpha$ (eIF2$\alpha$) is generally regarded as the canonical trigger of stress granule assembly: the resulting block in translation initiation drives polysome disassembly, and the released mRNA condenses with RNA-binding proteins and translation initiation complexes into stress granules \cite{kedersha1999stress, hofmann2021molecular}. Physically, stress granules form through liquid-liquid phase separation driven by multivalent interactions between RNA-binding proteins and mRNAs, generating dynamic, liquid-like structures that can be disassembled when stress is relieved \cite{guillen2020rna}. Stress granule formation and composition are also influenced by post-translational modifications of associated RNA-binding proteins, such as ubiquitination and acetylation \cite{wang2023post}. Given their role in cellular homeostasis, stress granules are considered protective against cellular damage, while their dysregulation contributes to pathology.

Alongside stress granules, phase separated processing bodies, commonly known as P-bodies, also play a key role in regulating mRNA, including their storage, degradation, and translational repression (Fig.\ref{fig:intro}c). These membraneless organelles are composed of mRNA decay enzymes, RNA-binding proteins, and other molecules that facilitate the turnover of mRNA, helping cells regulate gene expression in response to various stimuli \cite{parker2007p,decker2012p}. They also contain decapping enzymes (Dcp1/Dcp2) and exonucleases \cite{franks2008control}, as well as components of the RNA-induced silencing complex (RISC) machinery, which play a role in RNA interference \cite{liu2005microrna}. Unlike stress granules that primarily store mRNA during stress, P-bodies degrade and recycle mRNA \cite{eulalio2007p}. Despite this difference, P-bodies and stress granules frequently interact and share many protein and mRNA components; mRNPs within P-bodies can be remodeled and subsequently accumulate in stress granules, suggesting a dynamic cycle of mRNA exchange between polysomes and these compartments \cite{buchan2009eukaryotic}. Similar to many phase-separated structures, the dynamic nature of P-bodies permits selective processing of mRNAs based on cellular needs \cite{aizer2008dynamics}, an example of the partition-driven recruitment that Section~\ref{sec:Transport} treats formally.

The nucleolus is another key membraneless condensate that transcribes, processes, and assembles ribosomal RNA (rRNA) into ribosomes \cite{boisvert2007multifunctional}. Located within the nucleus, this phase separated dynamic structure is composed of three morphologically distinct structures, the fibrillar centers (FC), the dense fibrillar component (DFC) and the granular component (GC) -- three coexisting liquid phases whose nested architecture is a canonical example of multi-component coexistence that goes beyond the single-polymer Flory-Huggins theory \cite{feric2016coexisting,jaberi2023evolutionarily}. The nucleolus is also highly sensitive to cellular stress; changes in nutrient availability, environmental conditions, or cellular damage can cause nucleolar reorganization and disrupt its function \cite{boulon2010nucleolus}. This reorganization reflects the liquid-like properties of nucleolar components, which enable the nucleolus to act as a stress sensor by rapidly assembling and disassembling its biomolecular contents in response to environmental cues \cite{olson2004sensing,brangwynne2011active}.

Many additional condensates exist throughout the cell. For example, the TIS granule forms a meshwork intertwined with the endoplasmic reticulum and regulates translation, the process through which mRNA is converted into protein \cite{ma2018membraneless}. P granules reside in germ cells and regulate RNA and protein production associated with germline development and fertility \cite{wang2014p}. Neuronal RNA granules transport and localize mRNA to synapses, supporting synaptic plasticity \cite{krichevsky2001neuronal}. Finally, GW-bodies are known to sequester translationally repressed RNAs \cite{patel2016gw}. Together these examples illustrate the diversity of cellular locations and functions accessible to phase-separated structures.

\subsection{Two-dimensional and membrane-associated condensates}\label{sec:Types2D}
While the condensates described above are fully three-dimensional droplets dispersed through the cytoplasm or nucleoplasm, a growing family of phase-separated structures is tethered to the lipid bilayer (Fig.\ref{fig:intro}a). These membrane-associated two-dimensional condensates form when multivalent proteins recruited to a bilayer cluster into phase-separated patches \cite{banjade2014phase,snead2019control}. Membrane anchoring changes the physics in three concrete ways. First, the interfacial tension that bounds a three-dimensional droplet is replaced by a one-dimensional line tension, an energy per unit length, along the cluster boundary. Second, the dilute-to-dense phase transition can be triggered below the bulk saturation concentration through prewetting on the membrane surface, a mechanism that Cahn's classical analysis of critical-point wetting first laid out for binary fluid mixtures \cite{cahn1977critical}. Third, the elastic energy of the host bilayer, set by its bending modulus in the Helfrich sense, couples to cluster shape and material properties \cite{helfrich1973elastic,mangiarotti2025lipid}. The result is a class of condensates whose stability, geometry, and dynamics are governed by two-dimensional colloidal physics that the cytoplasmic droplets discussed above cannot illustrate.

The canonical example, depicted in Fig.\ref{fig:intro}d, is the T-cell receptor (TCR) signaling cluster. When the TCR/CD3 complex on a T cell engages an antigen held by a major histocompatibility complex (MHC) on an antigen-presenting cell, the transmembrane adaptor protein LAT (Linker for Activation of T cells) becomes phosphorylated by the kinase ZAP-70. LAT then recruits a multivalent network of adaptors and effectors, including Grb2, SOS, and the actin nucleation machinery. Su and colleagues showed that this network undergoes phase separation in the plane of the inner leaflet, forming submicron condensates that concentrate signaling components and exchange material on a timescale of seconds \cite{su2016phase}. Physically, the condensates reshape the local kinase-to-phosphatase ratio, excluding CD45 while retaining ZAP-70 so that LAT phosphorylation is sustained and downstream activation is enhanced relative to the same molecules dispersed on the membrane. The TCR cluster therefore exemplifies how a two-dimensional condensate emerging via phase separation physics can operate as a dynamic signal-processing element. 

Three further classes generalize the TCR picture. The first, historically the earliest, is the nephrin/Nck/N-WASP system at the kidney podocyte slit diaphragm. Phosphorylated nephrin tails on the inner leaflet recruit Nck, which in turn recruits N-WASP through multivalent interactions between Src-homology 3 (SH3) domains and proline-rich motifs, and the assembly phase separates \cite{li2012phase,banjade2014phase}. This was the system in which multivalent interactions were first shown to drive biomolecular phase separation, later formalized as sticker-and-spacer behavior \cite{choi2020physical}, and its reconstitution on supported bilayers remains a benchmark for membrane-associated condensates. The second is the postsynaptic density, in which scaffold proteins such as PSD-95, GKAP, SynGAP, and Shank co-assemble into a dense, phase-separated platform on the postsynaptic membrane and concentrate glutamate-receptor regulators within it \cite{zeng2018reconstituted}. The third is the tight junction, where the zonula occludens proteins ZO-1 and ZO-2 phase separate at cell-cell contacts to template the assembly of claudin- and occludin-based seal strands \cite{beutel2019phase}. Across all three, the unifying biophysical advantage is that assembly of multivalent scaffolds on membranes lowers the effective concentration threshold for phase separation relative to that in the cytosol, thus enabling localized biochemical processes that would otherwise require higher cytosolic concentrations.

Recent biophysical work has explored a related system - three-dimensional condensates associating with membranes. In cells, such membrane-condensate interactions underlie a growing list of processes, the formation and association of protein granules at the endoplasmic reticulum \cite{ma2018membraneless}, the assembly of endocytic vesicles \cite{day2021liquid}, and the plugging of damaged endolysosomes by stress granules to enable membrane repair \cite{bussi2023stress}. Reconstitutions in giant unilamellar vesicles show that the wetting and partitioning dynamics play a critical role in membrane-condensate interactions, which in turn are tightly controlled by the composition and material properties of the bilayer \cite{mangiarotti2025lipid}. In subsequent sections, we will comprehensively view condensates through the lens of colloidal and interfacial science to better understand their nature and physical interactions (Section~\ref{sec:Colloidal}), the origin of condensate pathologies (Section~\ref{sec:Pathologies}), and emerging strategies to develop and deliver drugs to condensates (Section~\ref{sec:Therapeutic}). 

\section{Colloidal and interfacial phenomena in condensates}\label{sec:Colloidal}
Viewed through the lens of colloidal and interfacial science, biomolecular condensates are dense liquid phases that have separated from their surroundings, driven by weak, multivalent interactions among biopolymers such as proteins and nucleic acids. Analogous to conventional phase separated systems, biomolecular condensates have been characterized through a large set of measurables: interfacial tension $\gamma$ \cite{brangwynne2011active,jawerth2018salt}, zero-shear viscosity $\eta_0$ \cite{elbaum2015disordered}, plateau modulus $G_N^0$ and viscoelastic relaxation time $\tau_R$ \cite{jawerth2020protein,alshareedah2024determinants}, partition coefficient $K$ \cite{wei2017phase,klein2020partitioning}, Flory-Huggins interaction parameter $\chi$, saturation and dense-phase concentrations $c_{\rm sat}$ and $c_{\rm dense}$ \cite{wang2018molecular}, diffusivity $D$ \cite{taylor2019quantifying,hubatsch2021quantitative}, interphase electrostatic potential, and membrane contact angles. Most of these are not independent. $c_{\rm sat}$ and $c_{\rm dense}$ can be linked to $\chi$, while rheological models such as the Maxwell model can relate $\eta_0$ to $G_N^0$ and $\tau_R$, and the generalized Stokes-Einstein can tie $D$ to $\eta_0$. Based on this and their relevance in condensate pathologies and therapeutic efforts, in this manuscript, we will focus on four properties to organize the physical behavior of condensates: the Flory-Huggins interaction parameter $\chi$, interfacial tension $\gamma$, the partition coefficient $K$, and the viscoelastic relaxation time $\tau_R$. This section and section \ref{sec:Transport} collectively describe each in detail with explicit scales and the measurement techniques that can quantify them. Section~\ref{sec:Pathologies} then interprets condensate pathologies as a perturbation of $\chi$, $\gamma$, $K$, and $\tau_R$ out of their functional ranges.

\begin{figure*}[!ht]
\vspace{-0.75 cm}
    \centering
    \includegraphics[width=0.9\linewidth]{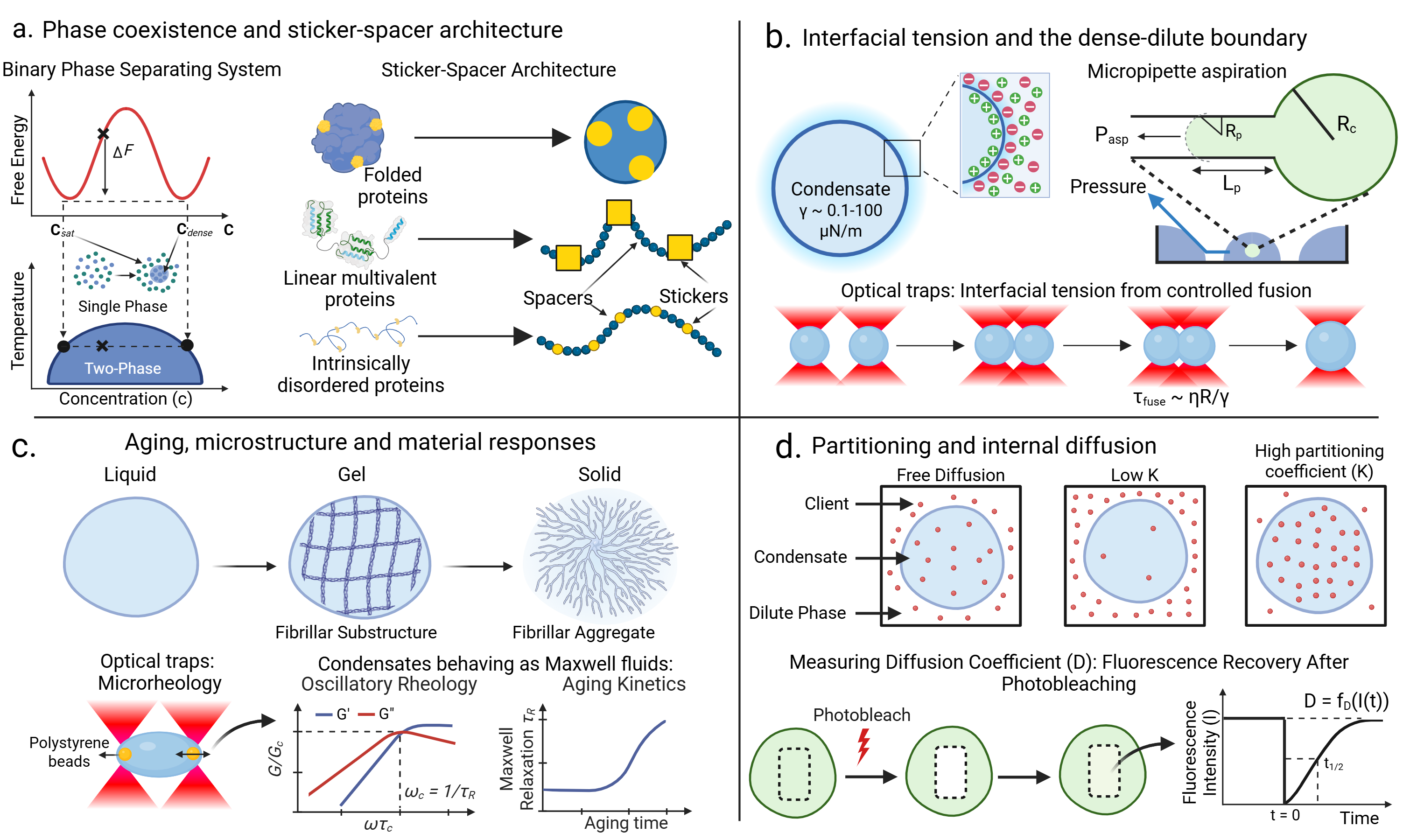}
    \caption{Colloidal, interfacial, and transport properties of condensates. {\bf a.} Phase coexistence and sticker-spacer architecture. When the interaction parameter $\chi$ exceeds its critical value, a homogeneous solution lowers its free energy by demixing into a dilute phase at $c_{\rm sat}$ and a dense phase at $c_{\rm dense}$. In folded, linear multivalent, and intrinsically disordered proteins, phase separation is driven by multivalent interaction motifs (stickers) connected by flexible segments (spacers). {\bf b.} Interfacial tension and the dense-dilute boundary. Condensates have a diffuse interface with a weak interfacial tension $\gamma$ and an electric double layer. $\gamma$ can be measured by micropipette aspiration or from the timescale of controlled fusion between optically trapped droplets, $\tau_{\rm fuse} \sim \eta_0 R/\gamma$. {\bf c.} Aging, microstructure, and material response. Condensates can age from a liquid to a gel with fibrillar substructure and finally to a solid fibrillar aggregate. Optical-trap microrheology is a common tool used to measure the storage and loss moduli of condensates. Many condensates behave as Maxwell fluids with a $\tau_R$ that grows as the condensate ages. {\bf d.} Partitioning and internal diffusion. A client molecule distributes between the dense and dilute phases according to its partition coefficient $K$, and its diffusion coefficient $D$ inside the condensate can be obtained from Fluorescence Recovery After Photobleaching (FRAP).}
    \label{fig:ColloidalInterfacial}
\end{figure*}

The thermodynamic foundation behind condensate formation is the Flory-Huggins lattice model for polymer-solvent phase coexistence, where the free energy of mixing per lattice site $f$ is
\begin{equation}
\frac{f(\phi_p)}{k_B T} = \frac{\phi_p}{N_p}\ln\phi_p + (1-\phi_p)\ln(1-\phi_p) + \chi\,\phi_p\,(1-\phi_p),
\label{eq:fh}
\end{equation}
with $\phi_p$ the polymer volume fraction, $N_p$ the number of monomers per chain, and $\chi$ the polymer-solvent interaction parameter \cite{flory1942thermodynamics,huggins1942theory, brangwynne2015polymer, zwicker2025physics}. When $\chi$ exceeds a critical value $\chi_c = \tfrac{1}{2}(1+1/\sqrt{N_p})^2$, the homogeneous solution lowers its total free energy by demixing: net effective attraction between polymers outweighs the total entropic cost of concentrating the polymers, and the system separates into a dilute phase with concentration $c_{\rm sat}$ and a condensed phase with $c_{\rm dense}$ (Fig.~\ref{fig:ColloidalInterfacial}a). For biopolymers that drive condensate formation, a large effective $\chi$ usually arises from weak, multivalent interactions rather than from a single chemistry. The interacting units are often organized in a sticker-and-spacer architecture (Fig.~\ref{fig:ColloidalInterfacial}a), such as in intrinsically disordered proteins, where short multivalent interaction motifs (stickers) are connected by flexible linker segments (spacers) \cite{choi2020physical}. The sticker chemistries include aromatic side chains, charged patches, polar and hydrophobic residues, and short folded motifs. Wang and colleagues showed that the saturation concentration of FUS-family proteins depends quantitatively on tyrosine and arginine sticker count, mechanistically linking $\chi$ to sequence chemistry \cite{wang2018molecular}. 

The cellular interior, with a multitude of macromolecules collectively having concentrations of 200--400~g/L, exerts excluded-volume effects that lower $c_{\rm sat}$ of phase separating biomolecules \cite{zimmerman1993macromolecular, ellis2001macromolecular}. Within the Flory--Huggins framework, excluded-volume effects imparted by crowders (ambient macromolecules) decrease the entropy of mixing, thus reducing $c_{\rm sat}$ of phase-separating molecules and enhancing condensate formation. Physiologically, cells often take advantage of crowding, for example, via altering ribosome concentration, to control condensate formation\cite{delarue2018mtorc1}. The same physics also plays a role in the concentration-dysregulation mechanism of condensate pathologies developed in subsequent sections.

Post phase separation, the dense phase (condensate) is bounded by a surface with an interfacial tension $\gamma$ (Fig. \ref{fig:ColloidalInterfacial}b). Physically, $\gamma$ is the excess free energy per unit area required to create an interface, and is directly related to free-energy density $f(\phi_p)/v_s$ under the Cahn-Hilliard square-gradient theory as \cite{cahn1958free, zwicker2025physics} 
\begin{equation}
\gamma = \int_{\phi_p^{\rm sat}}^{\phi_p^{\rm dense}} \sqrt{2\kappa_{\rm CH}\,\Delta f(\phi_p)/v_s}\,\,d\phi_p,
\label{eq:CahnHilliard}
\end{equation}
where $\kappa_{\rm CH}$ is the gradient stiffness associated with composition variations across the interface, $v_s$ is lattice-site volume, and $\phi_p^{\rm sat}$, $\phi_p^{\rm dense}$ are the polymer volume fractions of the coexisting dilute and dense phases, respectively. Here $\Delta f(\phi_p) = f(\phi_p) - f_{\rm coex}(\phi_p)$ is the local free-energy penalty for holding material at an intermediate composition $\phi_p$ rather than demixing it into the two coexisting phases, where $f_{\rm coex}(\phi_p)$ is the common-tangent free energy of the demixed system at $\phi_p$. For biomolecular condensates, measured values span $\gamma \sim 0.1$--$100~\mu$N/m, orders of magnitude below the $\sim 72$~mN/m of an air-water interface \cite{brangwynne2009germline, brangwynne2011active,jawerth2018salt,elbaum2015disordered,holland2023surface,law2023bending}. Two consequences follow. First, the Laplace pressure $\Delta P = 2\gamma/R$ across a condensate of radius $R \sim 1~\mu$m is of order pascals, two to five orders of magnitude weaker than across an emulsion droplet of the same size \cite{wang2021surface}. Condensates are therefore easily deformed by cytoskeletal contacts, by fluid flow, and by membranes that wet them. Second, low $\gamma$ leads to long capillary timescales $\tau_{\rm fuse} \sim \eta_0 R / \gamma$, resulting in retarded condensate fusion. In vivo, condensate fusion is further counteracted in some systems by surface-active proteins that adsorb to the condensate interface with desorption energies far exceeding $k_BT$, sterically arresting condensate fusion in a mechanism analogous to Pickering stabilization of emulsions. MEG-3 clusters on PGL-3 condensates in \textit{C.~elegans} are canonical examples of Pickering stabilization of condensates \cite{folkmann2021regulation}. Interfaces of condensates can carry an electric double layer depending on the charged residues in phase separating biopolymers, with reported positive and negative interphase potentials in the range of millivolts to tens of millivolts \cite{posey2024biomolecular, dai2023interface}, which in turn can drive spontaneous redox chemistry at the interface \cite{dai2023interface}. Electrostatic contributions to $\chi$ also make $\gamma$ sensitive to protein charge and ambient ionic strength \cite{jawerth2018salt, spruijt2009interfacial, holland2023surface}. 

$\gamma$ is commonly obtained by measuring the capillary timescales of condensate fusion, breakup, or shape relaxation events \cite{brangwynne2011active, jawerth2018salt}. As shown in Fig.~\ref{fig:ColloidalInterfacial}b, complementary methods include micropipette aspiration, which directly imposes a pressure differential and reads off $\gamma$ from the equilibrium mean curvatures of the condensate droplet and the liquid interface inside the micropipette \cite{wang2021surface}; thermal capillary-wave (flicker) spectroscopy of the dense-dilute boundary, which extracts $\gamma$ from the spontaneous fluctuation spectrum \cite{ng2025microscopy, jawerth2018salt}; sessile drop tensiometry, which obtains $\gamma$ from the shape of a sessile drop\cite{holland2023surface}, and atomic-force microscopy with a colloidal probe, which yields $\gamma$ from force curves during probe retraction \cite{santamaria2024frapid}.

A condensate's interior is rarely a simple Newtonian fluid. Multivalent cross-linking, especially among intrinsically disordered regions, generates viscoelastic materials which deform elastically at short timescales and viscously at longer times (Fig.~\ref{fig:ColloidalInterfacial}c). The simplest rheological representation of a viscoelastic condensate is via a Maxwell element, where the complex modulus goes as
\begin{equation}
G^*(\omega) = G_N^0 \frac{i\omega\tau_R}{1 + i\omega\tau_R}.
\label{eq:maxwell}
\end{equation}
Here $G_N^0$ is the plateau elastic modulus and $\tau_R$ is the relaxation time at which storage and loss moduli cross; the zero-shear viscosity follows as $\eta_0 = G_N^0 \tau_R$ \cite{wang2025rheology}. For freshly assembled liquid condensates, $\tau_R$ is of order $10^{-3}$ to $10^0$~s and $\eta_0$ ranges from $0.01$ to $10$~Pa$\cdot$s, one to four orders of magnitude above water\cite{patel2015liquid,elbaum2015disordered,jawerth2018salt}. For aberrantly aged condensates, $\tau_R$ can grow by several orders of magnitude with the material approaching a gel or a soft glass \cite{jawerth2020protein,patel2015liquid, alshareedah2024sequence}. This process can occur via multiple documented routes: $\beta$-strand zippering between intrinsically disordered regions (IDRs), disulfide cross-linking of cysteine residues under oxidizing conditions \cite{poole2015basics,sies2017oxidative}, and slow nucleation of fibrillar substructure (Fig.~\ref{fig:ColloidalInterfacial}c), all acting as effective stickers that can lock $\tau_R$ at high values. This transition from a liquid that relaxes on accessible timescales to a solid-like network is the central physical signature of several condensate pathologies.

Two-dimensional and membrane-tethered condensates have colloidal physics parallels to the three-dimensional treatment above. In these cases, the Flory-Huggins model of phase separation can be cast as a surface coexistence in the membrane plane, with a surface saturation $\Gamma_{\rm sat}$ replacing $c_{\rm sat}$ \cite{banjade2014phase, zhao2025theory}. The interfacial tension $\gamma$ becomes a one-dimensional line tension $\lambda$ along the cluster boundary. A distinctive two-dimensional phenomenon is prewetting: a membrane that sits near a critical demixing point facilitates coexisting surface phases at bulk concentrations well below the phase separation threshold, expanding the regime over which surface phase coexistence is possible \cite{cahn1977critical,rouches2021surface}. Membrane recruitment can therefore promote condensation at substantially lower protein concentrations than required in solution, as shown on model membranes for Whi3 ribonucleoprotein condensates that are known to associate with the endoplasmic reticulum in cells \cite{snead2022membrane}. An intermediate system between purely two-dimensional and purely three-dimensional condensates arises when a condensate in the bulk interacts with a lipid bilayer\cite{mangiarotti2024biomolecular}. The interacting condensate adopts a contact angle $\theta$ set by Young's balance of the three interfacial tensions (condensate--dilute, condensate--membrane, and membrane--dilute)\cite{mangiarotti2024biomolecular, mangiarotti2025lipid}, with the membrane contribution modified by the Helfrich bending energy of the bilayer \cite{helfrich1973elastic}. The lipid composition and packing matter: cholesterol content and acyl-chain packing tune both the wetting energy and the partitioning of the condensate between liquid-ordered (Lo) and liquid-disordered (Ld) phases on the bilayer \cite{mangiarotti2025lipid}.

\section{Transport phenomena in condensates}\label{sec:Transport}
Transport in and out of biomolecular condensates spans three distinct regimes. Inside the dense phase, individual molecules diffuse through a crowded, viscoelastic medium. At the dense-dilute boundary, client molecules cross an interface whose chemistry contributes to a partition coefficient and a finite mass-transfer resistance (Fig.~\ref{fig:ColloidalInterfacial}d). Beyond the boundary, condensates themselves can move on cytoplasmic length scales, driven by concentration gradients, by Marangoni stresses, or by ATP-consuming active machinery. Each regime carries pathological signatures as well: aged condensates lose diffusive mobility, dysregulated post-translational modifications alter partition coefficients, and loss of active driving collapses spatial organization \cite{niu2023biomolecular,alshareedah2024determinants,berry2018physical,ryan2021tyrosine}.

The basic partitioning relation follows from chemical-potential equality of a client species that can move between the dense and dilute phases. For a client at low concentration,
\begin{equation}
K = \frac{c_{\rm in}}{c_{\rm out}} = \exp\!\left(-\frac{\Delta G_{\rm part}}{k_B T}\right),
\label{eq:Kpart}
\end{equation}
with $\Delta G_{\rm part}$ the free-energy difference of inserting one client molecule into the dense versus the dilute phase, set by both enthalpic interactions with the dense-phase polymers and entropic excluded-volume penalties. Experimentally, $K$ is measured as the ratio of client fluorescence inside and outside droplets by confocal microscopy \cite{klein2020partitioning}, from fluorescence correlation spectroscopy (FCS)-determined concentrations in each phase \cite{wei2017phase}, or as a fit parameter from spatially resolved fluorescence recovery after photobleaching (FRAP) without requiring dilute-phase concentration measurements \cite{hubatsch2021quantitative}. Reported values of $K$ for biological clients span $10^{-1}$ to $10^3$: some species are nearly excluded, while others are concentrated by three orders of magnitude relative to bulk \cite{wei2017phase,klein2020partitioning}. $K$ is tuned by the physical characteristics of the client molecule (e.g., charge, aromaticity, hydrophobicity) and by post-translational modifications such as phosphorylation that can alter a residue's effective interaction with the dense-phase scaffold \cite{klein2020partitioning, emelianova2025prediction, wessen2025sequence, wang2018molecular,guo2019pol,ryan2021tyrosine}. Because $\Delta G_{\rm part}$ enters Eq.~\ref{eq:Kpart} exponentially, a 1--2~$k_B T$ shift redistributes a client by close to an order of magnitude, the quantitative basis for the partition-dysregulation mechanism of Section~\ref{sec:Pathologies}.

Once inside a condensate, a client molecule is transported primarily by diffusion, which for viscoelastic condensates can be described by the generalized Stokes-Einstein relation
\begin{equation}
D(\omega) = \frac{k_B T}{6\pi a\,\eta^*(\omega)},
\label{eq:GSE}
\end{equation}
where $a$ is the hydrodynamic radius of the client and $\eta^*(\omega)$ is the complex viscosity \cite{mason1995optical}, which is related to the complex modulus $G^*(\omega)$ as $\eta^*(\omega) = G^*(\omega)/(i\omega)$. For purely viscous condensates, or for timescales much longer than the relaxation time ($\omega\tau_R \to 0$), $\eta^*(\omega)$ reduces to the zero-shear viscosity $\eta_0$, and Eq.~\ref{eq:GSE} reduces to the classical Stokes-Einstein form $D = k_B T/(6\pi \eta_0 a)$. The classical form is also recovered in the limit of clients becoming smaller than the mesh size of the condensate polymeric network, with diffusing clients experiencing a local viscosity that can be smaller than $\eta_0$ \cite{wei2017phase}. As is evident from Eq.~\ref{eq:GSE}, condensate viscosity (or in general complex viscosity) has an inverse relationship with client mobility, meaning an abnormal increase in condensate viscosity can hamper the biological activity of client molecules and is a major source of condensate pathologies. Measured values of $D$ in condensates span $10^{-3}$ to $10\; {\mu}m^2/s$ \cite{hubatsch2021quantitative,taylor2019quantifying}. As shown in Fig.~\ref{fig:ColloidalInterfacial}d, the diffusion coefficient in condensates is most commonly measured via FRAP, a non-invasive method where fluorescence intensity changes in photobleached spots on a condensate are tracked over time and fitted to a model to recover diffusion coefficients \cite{hubatsch2021quantitative,taylor2019quantifying}. Fluorescence correlation spectroscopy, raster image correlation spectroscopy and single-particle tracking are additional non-invasive techniques that can be used to obtain $D$ \cite{wei2017phase, ng2025microscopy, visser2025controlling}, but like FRAP, do not resolve the full viscoelastic spectrum \cite{wang2025rheology}. Passive and active microrheology techniques can measure the frequency-dependent moduli $G^*(\omega)$, but require introducing tracer beads into condensates or deforming them under calibrated optical traps \cite{jawerth2020protein,alshareedah2024determinants}.

Crossing the dense-dilute interface is a kinetic process distinct from internal diffusion. The interface can impart a mass-transfer resistance, quantified by a mass-transfer Biot number $\mathrm{Bi} = \kappa R / D$, where $\kappa$ is the interface conductance (units m/s), $R$ is the condensate radius, and $D$ the dense-phase diffusivity. Mechanistically, FRAP recovery experiments have found that low interfacial conductance originates from the inability of molecules to cross the condensate in a single encounter, which in turn depends on the physical interactions between the client and condensate molecules (the fewer the free stickers on the client, the lower the interfacial conductance) \cite{zhang2024exchange}. The interface can also contribute to the bulk transport of condensates via Marangoni stresses originating from gradients in interfacial tension, commonly induced by concentration gradients of solutes in the cytosol \cite{jambon2024phase}. An interfacial-tension contrast $\Delta\gamma \sim 1~\mu$N/m across a condensate of viscosity $\eta \sim 1$~Pa$\cdot$s drives a Marangoni velocity $v \sim \Delta\gamma/\eta \sim 1~\mu$m/s, fast enough to matter on the cellular time- and length-scales. Concentration gradients of solutes can also drive bulk translation via diffusiophoresis, the migration of colloidal particles along solute concentration gradients \cite{annunziata2024diffusiophoresis,doan2024diffusiophoresis}.

Many condensates are not at equilibrium but are continuously driven by ATP-consuming reactions: P granules undergo position-dependent dissolution and condensation driven by MEX-5 gradients originating from kinase/phosphatase cycles \cite{saha2016polar,griffin2011regulation}, stress granules and nucleoli maintain their liquid-like dynamics by continuous ATP-dependent processes \cite{protter2016principles, brangwynne2011active}, and transcriptional condensate compositions and dynamics are driven by processes that require ATP \cite{guo2019pol, sabari2018coactivator}. These active-liquid-like condensates can yield behavior with no equilibrium analogs, such as suppressed Ostwald ripening \cite{zwicker2015suppression} and phase compositions set by non-equilibrium fluxes at the interface \cite{cho2025interfacial}. Loss of active driving is yet another route to the pathological liquid-to-solid transitions, and an emerging frontier for therapeutic interventions.

\section{Biomolecular condensates in pathologies}\label{sec:Pathologies}
\begin{figure*}[!ht]
\vspace{-0.75 cm}
    \centering
    \includegraphics[width=0.9\linewidth]{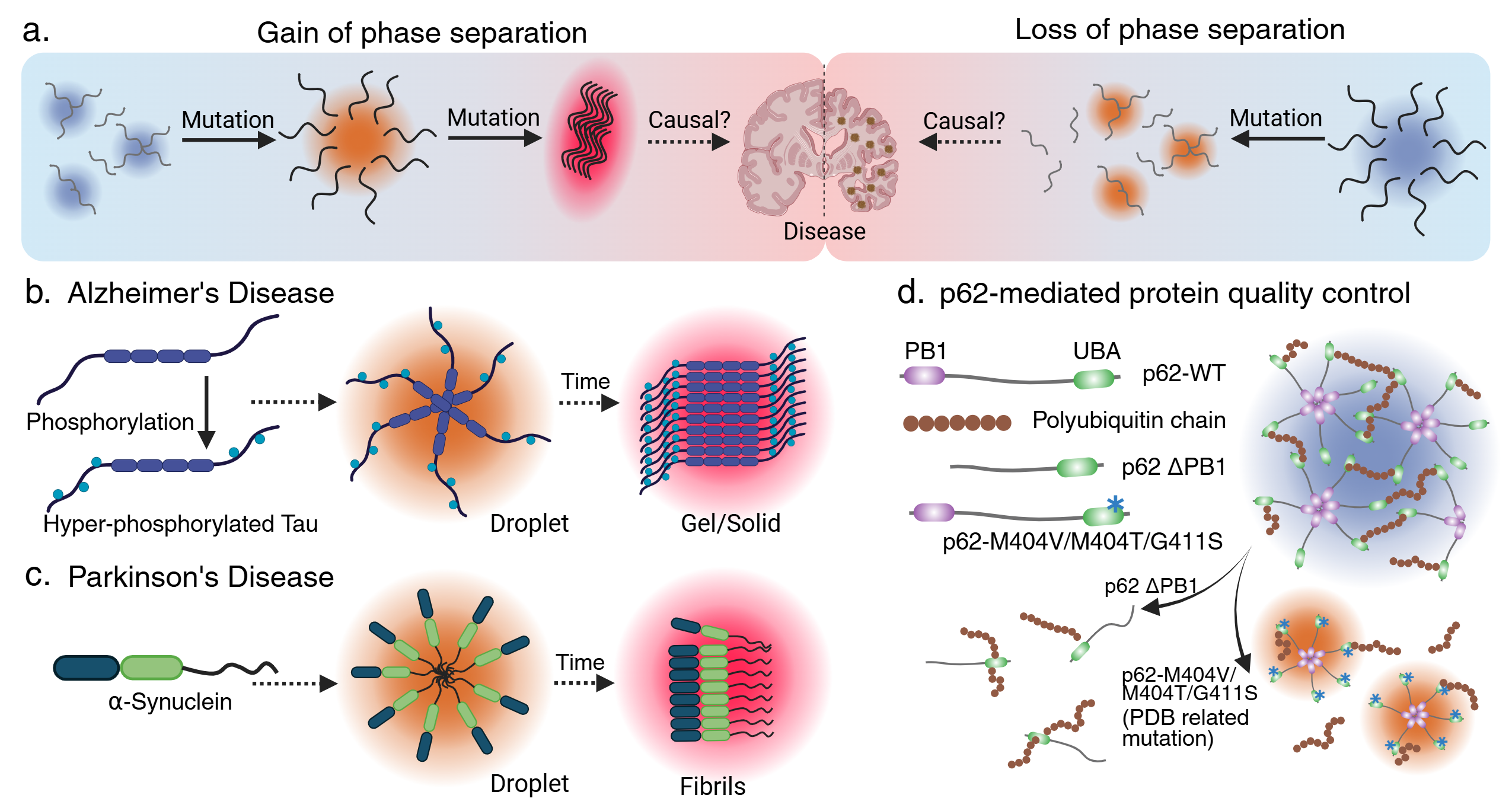}
    \caption{Physical failure modes in condensate pathologies. {\bf a.} Mutations can shift the saturation concentration in either direction: a gain of phase separation drives proteins to pathologically condense and subsequently solidify where they should not, while a loss of phase separation dissolves condensates that are required for function. Whether condensate pathologies by themselves are causal for disease remains an open question in many conditions. {\bf b.} In Alzheimer's disease, hyperphosphorylated tau condenses into droplets that age over time into a gel or solid. {\bf c.} In Parkinson's disease, $\alpha$-synuclein at pathological concentrations condenses into droplets that nucleate fibrils from within. {\bf d.} In Paget's disease of bone (PDB), UBA-domain mutations of p62 (M404V, M404T, and G411S) impair condensate formation. Wild-type p62 polymerizes through its PB1 domain and binds polyubiquitin chains through its UBA domain; deleting PB1 or mutating UBA reduces the multivalency that drives condensation.}
    \label{fig:pathologies}
\end{figure*}

Diseases of biomolecular condensates can be understood as failures of colloidal and transport properties of condensates or condensate-associated biomolecules leading to abnormal phase separation and condensate dynamics (Fig.~\ref{fig:pathologies}). This section organizes the condensate pathology literature into five mechanism classes: (a) saturation concentration dysregulation, (b) aberrant aging and liquid-to-solid transitions,
(c) disrupted partitioning and recruitment, (d) interfacial aggregation, and (e) pathogen hijacking, each defined by which physical principle of Eqs.~\ref{eq:fh}--\ref{eq:GSE} fails and how. 

\subsection{Saturation concentration dysregulation}\label{sec:PathCrowding}
A major class of condensate pathologies are associated with the levers that control phase separation itself (Eq.~\ref{eq:fh}): mutations or expression-level changes shift the saturation concentration $c_{\rm sat}$ or equivalently the polymer-solvent compatibility $\chi$ such that phase separation occurs where it should not, or fails to occur where it should (Fig.~\ref{fig:pathologies}a).

Cataract provides a clear example. The transparent eye lens contains an extremely concentrated $\gamma$-crystallin solution in a single, spatially uniform phase \cite{benedek1997cataract}. Increasing the inter-protein attraction, an effective $\chi$ shift arising from a point mutation, oxidation, or aging, shifts the phase-separation boundary into the physiological concentration range \cite{benedek1997cataract, stradner2007insight}. The resulting demixing and condensate formation produces spatial variations in refractive index, which scatters light, and makes the lens opaque. Cancer driven by mutations in speckle-type POZ protein (SPOP) illustrates the opposite failure: condensation that does not occur where it should. SPOP and its substrates trigger phase separation into nuclear bodies whose assembly is required for full ubiquitination activity in cells \cite{bouchard2018cancer}. Tumor-associated mutations in the substrate-binding domain weaken SPOP-substrate interaction, effectively raising $c_{\rm sat}$ (or equivalently lowering $\chi$) such that mutant SPOP fails to co-condense with its substrates, leading to oncogenic clients escaping degradation and losing the tumor-suppressor function \cite{bouchard2018cancer}.

Several pathologies discussed in the subsequent subsections also have their origin in saturation concentration dysregulation. Tau exists at $\sim 2~\mu$M in neurons, near its measured saturation concentration for LLPS; hyperphosphorylation lowers the effective $c_{\rm sat}$ and pushes tau across the binodal --- a process that is proposed as starting the condensate-to-fibril cascade of Alzheimer's disease (Fig.~\ref{fig:pathologies}b) \cite{wegmann2018tau,zbinden2020phase}. $\alpha$-Synuclein in Parkinson's disease provides a complementary example: crowding and cellular overexpression, including familial cases of \emph{SNCA} gene duplication and triplication \cite{miller2004alpha}, raise $\alpha$-synuclein above its effective $c_{\rm sat}$ and initiates the LLPS-to-fibril trajectory (Fig.~\ref{fig:pathologies}c) \cite{ray2020alpha}. Conversely, in Paget's disease of bone (Fig.~\ref{fig:pathologies}d), UBA-domain mutations of p62 impair the multivalent UBA-polyubiquitin interactions that drive condensate formation, effectively raising the $c_{\rm sat}$ for polyubiquitin-induced phase separation \cite{sun2018polyubiquitin}. In both directions, the aberrant phase separation is controlled by the biomolecular concentration relative to $c_{\rm sat}$ or the effective $\chi$; the subsequent material-state and partitioning failures associated with these condensate pathologies are treated in Sections~\ref{sec:PathAgeing} and \ref{sec:PathPartition}.

\subsection{Aberrant aging and liquid-to-solid transitions}\label{sec:PathAgeing}
The most studied class of condensate pathology is the liquid-to-solid transition of functional condensates into pathological gels, fibrils, or amyloids. In healthy condensates, the viscoelastic relaxation time $\tau_R$ stays short enough that condensates behave as viscous liquids, allowing diffusing clients to sample the interior rapidly. Under disease-associated mutations or post-translational modifications, the effective sticker chemistry changes: $\beta$-strand zippering between intrinsically disordered regions, hydrogen-bonded amyloid contacts, and accumulated misfolding all produce effectively permanent stickers that drive solid-like behavior in condensates with $\tau_R$ increasing by orders of magnitude \cite{jawerth2020protein, alshareedah2024sequence}. As expected from Eq.~\ref{eq:GSE}, internal viscosity also rises and client diffusion stalls; the condensate loses its biochemical function.

Several protein systems are commonly associated with this pathological mechanism. Tau, normally a microtubule-associated regulator, phase separates into liquid droplets that age into gels and ultimately into aggregates characteristic of Alzheimer's disease (Fig.~\ref{fig:pathologies}b); hyperphosphorylation accelerates this transition by enabling effectively irreversible sticker interactions (attributed to exposed hydrophobic hexapeptides) between tau monomers \cite{zbinden2020phase,nedelsky2019bridging,wegmann2018tau}. $\alpha$-Synuclein in Parkinson's disease follows a parallel trajectory: monomeric protein concentrates into a metastable droplet which nucleates fibril growth from within (Fig.~\ref{fig:pathologies}c), with the LLPS step now proposed as a nucleation precursor to fibril formation \cite{ray2020alpha}. FUS and TDP-43, both RNA-binding proteins implicated in amyotrophic lateral sclerosis (ALS) and frontotemporal dementia (FTD), undergo analogous liquid-to-solid transitions whose kinetics are accelerated by disease-associated mutations such as FUS G156E and TDP-43 M337V \cite{patel2015liquid,nedelsky2022pathological, yan2025intra}. Stress granules, while protective \cite{glineburg2024stress}, can turn pathological when misfolded proteins partition into them, raising $\tau_R$ and driving a liquid-to-solid transition within the granule \cite{mateju2017aberrant,patel2015liquid}. 

Liquid-to-solid transitions in condensates can also be driven by the state of the cytosol rather than the chronic aging of the protein itself \cite{yan2025intra}. Cysteine residues in condensate-forming proteins are redox-sensitive \cite{poole2015basics}, and under oxidative stress they form intermolecular disulfide bonds that act as stickers \cite{yan2025intra}. These sticker interactions can facilitate cross-$\beta$ sheet formation between proteins, driving irreversible liquid-to-solid transitions characteristic of pathological aggregates. This redox channel is commonly associated with neurodegeneration, where mitochondrial dysfunction and accumulated reactive oxygen species are linked to liquid-to-solid transitions in TDP-43 condensates \cite{sbodio2019redox, yan2025intra}.

\subsection{Disrupted partitioning and recruitment}\label{sec:PathPartition}
A third class of pathology emerges via disrupted partitioning of client molecules: mutations or post-translational modifications that change $\Delta G_{\rm part}$ for a client species can shift its partition coefficient $K$ by orders of magnitude, removing it from condensates where it is normally concentrated or recruiting it to condensates where it does not belong. Because $\Delta G_{\rm part}$ enters Eq.~\ref{eq:Kpart} exponentially, even a 1--2~$k_B T$ interaction-energy shift redistributes a client by a factor close to ten.

A direct example comes from HMGB1, a low-specificity DNA-binding protein that carries an acidic disordered C-terminal tail and is normally distributed through the nucleoplasm. De novo frameshift variants can replace this acidic tail with an arginine-rich basic one, resulting in the mutant protein pathologically partitioning into nucleoli and displacing the NPM1-rich granular component that normally occupies them \cite{mensah2023aberrant}. This enhanced partitioning is mediated by arginines introduced by the mutations, which effectively act as stickers that lower $\Delta G_{\rm part}$ and facilitate entry into the nucleolus \cite{mensah2023aberrant}. The mispartitioned protein arrests nucleolar dynamics, lowering ribosomal RNA production, nascent translation and cell viability --- a pathological cascade that is linked to diseases such as brachyphalangy, polydactyly and tibial aplasia syndrome \cite{mensah2023aberrant}. Analogously, in \emph{C9orf72} ALS/FTD, the arginine-rich dipeptide-repeat proteins poly-GR and poly-PR partition into compartments rich in RNA-binding proteins and proteins rich in low-complexity sequence domains, such as the nucleoli, stress granules, and nuclear pores, where they pathologically accumulate and disrupt their function \cite{lee2016c9orf72}. A similar gain-of-$K$ mechanism is thought to operate in cancer: oncogenic fusion proteins with disordered domains (e.g., EWS-FLI1 in Ewing sarcoma) drive aberrant transcriptional condensates at de novo enhancers that raise the effective $K$ for chromatin-remodeling complexes \cite{boulay2017cancer}.

\subsection{Interfacial aggregation}\label{sec:InterfacialAggregation}
Condensate interfaces can act as an origin for pathologies and can themselves become pathogenic, defining a $\gamma$-coupled (interface-mediated) disease channel distinct from the bulk pathologies. Condensate interfaces can promote and accelerate protein aggregation \cite{choi2024condensate}, such as in the case of fibrils formed at the dense-dilute boundary by the low-complexity domain of hnRNPA1, a protein involved in amyotrophic lateral sclerosis \cite{linsenmeier2023interface}. A related interfacial mechanism is documented for $\alpha$-synuclein associated with Parkinson's disease: $\alpha$-synuclein adsorbs reversibly into multiple layers at the condensate surface, with the protein's amphiphilicity and interfacial $\zeta$-potential governing the adsorption process and seeding heterogeneous nucleation that accelerates downstream fibril growth \cite{visser2025controlling}.

\subsection{Pathogen hijacking}\label{sec:PathHijack}
The final class highlighted in this review is not a failure of physiological condensate physics but its exploitation: pathogens carry IDR-rich proteins that either nucleate new condensates
inside host cells, producing a gain-of-phase-separation through an effective $\chi$ shift driven by viral RNA--protein interactions \cite{iserman2020genomic}, or partition into existing host condensates and reprogram them through a $K$ shift \cite{luo2021sars}.

RNA viruses provide the cleanest examples. The nucleocapsid (N) protein of SARS-CoV-2 phase separates with viral genomic RNA into condensates \cite{iserman2020genomic}, which have been proposed to locally concentrate viral RNA and associated factors and to support viral assembly \cite{martin2023biomolecular}. Influenza virus organizes its ribonucleoprotein complexes into similar membraneless compartments at endoplasmic reticulum exit sites \cite{etibor2021liquid, alenquer2019influenza}, and the host response to both viruses includes stress granule formation \cite{guan2023multiple}
 that entraps viral messenger RNA and restricts its translation \cite{mccormick2017translation}. Some viruses turn this competition to their advantage: SARS-CoV-2 N protein phase separates with G3BP1/2 to disassemble host stress granules and facilitate viral production \cite{luo2021sars}, while West Nile virus can recruit a subset of stress granule components (e.g., TIAR) and P-body components (e.g., LSM1, GW182, DDX3) to viral replication sites where they positively regulate viral replication \cite{chahar2013p,beckham2008p}. Nucleolar hijacking adds a third route. Multiple viruses target nucleolar proteins or sequester ribosome-biogenesis machinery in their replication compartments, producing nucleolar reorganization that has been linked to perturbation of host stress-response signaling and to substrate supply for the virus \cite{iarovaia2021manipulation,hiscox2007rna}.

\begin{figure*}[!ht]
\vspace{-0.75 cm}
\centering
    \includegraphics[width=\linewidth]{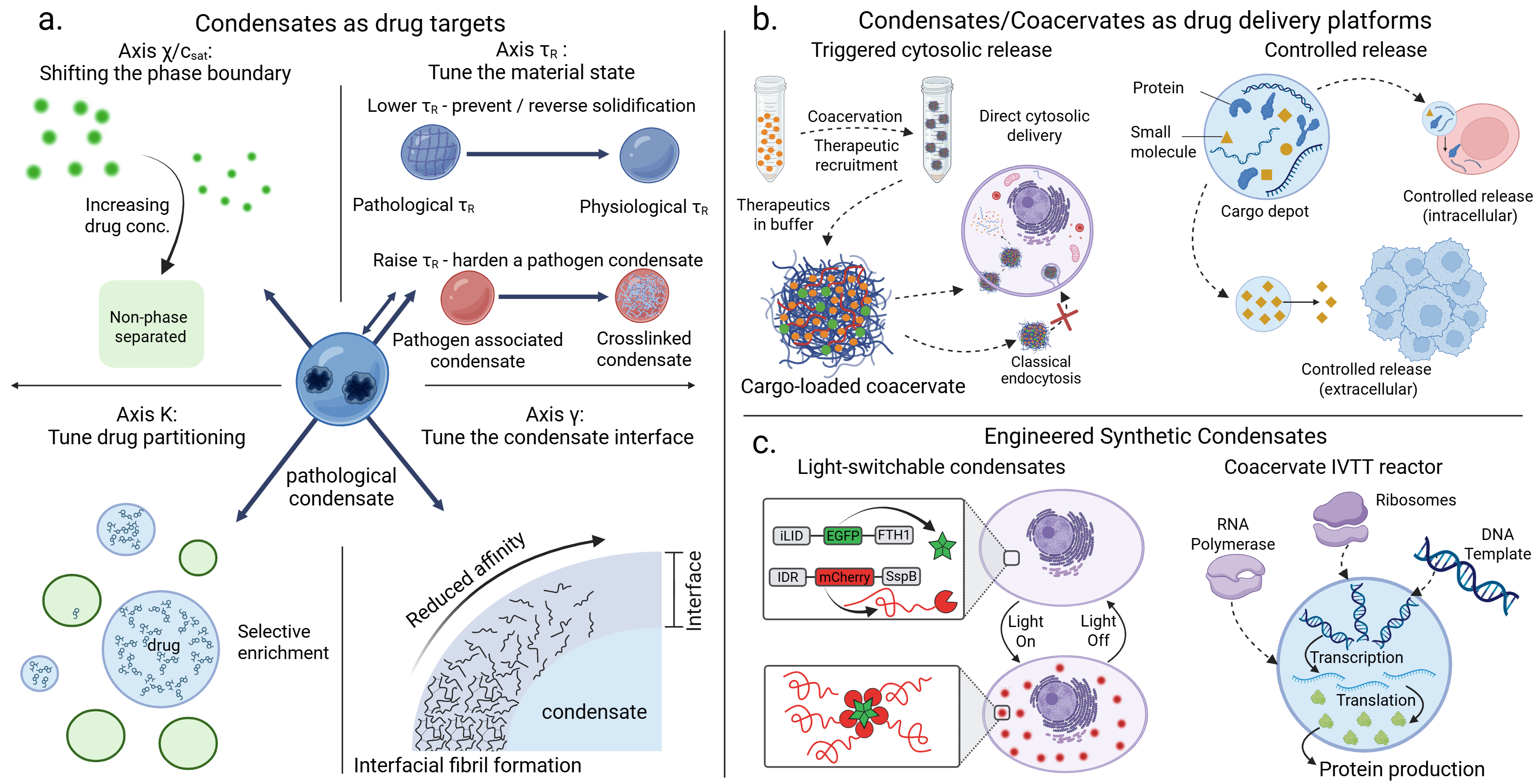}
    \caption{Therapeutic targeting and engineering of condensates. {\bf a.} A pathological condensate can be therapeutically targeted along the four physical axes discussed in the manuscript. Raising the concentration of a drug that competes for sticker contacts can shift the phase boundary through $\chi$ and $c_{\rm sat}$ until the system no longer phase separates. Lowering $\tau_R$ can reverse the solidification of an aged condensate, while raising $\tau_R$ can harden a pathogen-associated condensate and block its function. Tuning the chemistry of a drug to match the dense phase can raise its partition coefficient $K$ and selectively enrich it in the target condensate to increase therapeutic efficacy. Similarly, engineering drugs that can restore physiological client partitioning can also address partitioning-driven pathologies. Finally, engineering drugs that act at the interface ($\gamma$ axis) can reduce the affinity of amyloidogenic proteins for the dense-dilute boundary and suppress interface-mediated fibril formation.  {\bf b.} Coacervates as delivery vehicles. Therapeutics loaded into coacervates can facilitate effective drug delivery to the cytosol without classical endocytosis. Drug-loaded coacervates can also be used as a depot for controlled intracellular and extracellular release. {\bf c.} Engineered synthetic condensates. Light-switchable scaffolds can be used to assemble synthetic condensates in living cells on demand with desired valence and concentration. Synthetic coacervates are also enabling cell-free synthesis, such as in vitro transcription-translation (IVTT) coacervate reactors for producing proteins.}
    \label{fig:condensatetherapeutic}
\end{figure*}

\section{Therapeutic targeting and engineering of condensates}\label{sec:Therapeutic}

Across all five pathological mechanism classes of Section~\ref{sec:Pathologies}, the structure of the argument is the same: a physical quantity that defines functional condensate behavior is pushed outside its working range, and disease emerges. The design question for therapy can therefore be shifted from `which protein is involved?' to a biophysical one: `which physical process is failing, and which parameter must be restored?'. Pathological mechanism (a) suggests interventions that restore the working $c_{\rm sat}$ or shift the effective $\chi$ back toward the physiological range; mechanism (b) suggests interventions that lower the effective $\tau_R$ (small molecules that compete for sticker contacts; antioxidants that prevent cysteine cross-linking); mechanism (c) suggests interventions that re-tune $\Delta G_{\rm part}$ for the relevant client; mechanism (d) suggests interventions targeting the condensate interface (surfactant-like proteins or charge tuning); mechanism (e) suggests blocking the chemistry through which the pathogen modulates $\chi$ or $K$ (Fig.~\ref{fig:condensatetherapeutic}a). These design principles structure the therapeutic subsection on condensates as drug \emph{targets}: therapeutic molecules are engineered to enter, modify, or dissolve pathological condensates and restore the relevant parameter.

In a completely different vein, the second subsection focuses on the emerging research on condensates themselves as drug \emph{carriers}: engineered coacervates and synthetic phase-separating systems concentrate therapeutic payloads and release them on cue, connecting the modern biomolecular-condensate literature to classical colloidal drug delivery. Both of the disparate threads draw on the biophysical principles in Sections~\ref{sec:Colloidal} and~\ref{sec:Transport}, and both are under active research.

\subsection{Condensates as drug targets}\label{sec:TherapTargets}
Condensate-targeted therapies have advanced furthest for two of the pathological mechanism classes discussed in Section~\ref{sec:Pathologies}, namely, aberrant aging and liquid-to-solid transitions (mechanism (b)) and disrupted partitioning (mechanism (c)). For mechanism (b), the leading clinical candidate is anle138b (emrusolmin, TEV-56286), a small molecule that binds within cavities of $\alpha$-synuclein and amyloid-$\beta$ fibrils, modulating oligomerization and inhibiting fibril formation \cite{wagner2013anle138b,antonschmidt2022clinical,han2025anle138b}, an intervention that is aligned with the liquid-to-solid trajectory of mechanism (b). Anle138b has completed a Phase~1 trial in Parkinson's disease (ClinicalTrials.gov NCT04685265) and has advanced to a Phase~2 trial in multiple system atrophy (NCT06568237); see Table~\ref{tab:clinical} for details. Upstream of clinical development, ATP-mimetic hydrotropes have been proposed to dissolve aged condensates \cite{patel2017atp}, and Uechi and colleagues recently demonstrated that lipoamide, acting primarily through the redox-sensitive stress granule protein SFPQ, dissolves stress granules and rescues motor phenotypes in cellular and \textit{Drosophila} models of FUS- and TDP-43-driven ALS \cite{uechi2025small, wolozin2019stress}. 

For mechanism (c), the key evidence on condensate drug partitioning underlying current clinical and preclinical programs was established by Klein and colleagues, who showed that cisplatin partitions into MED1 transcriptional condensates with a partition coefficient up to $\sim$600 and that a fluorescent tamoxifen analog accumulates in estrogen-receptor-containing MED1 condensates, both with minimal accumulation in non-cognate condensates \cite{klein2020partitioning}. The leading clinical candidate exploiting this principle is DPTX3186 (Table \ref{tab:clinical}), a first-in-human Phase~1 compound for Wnt-pathway-activated solid tumors that began enrolling in 2026 under Dewpoint Therapeutics sponsorship (ClinicalTrials.gov NCT07312903). Dewpoint has also recently announced an oral MYC condensate modulator at the late-preclinical development-candidate stage \cite{ren2026oral}. Upstream, broader oncology programs target MED1, MYC, and FOXM1 transcriptional condensates, either with small molecules that selectively partition into them \cite{klein2020partitioning} or with small molecules that destabilize the IDR contacts within condensates \cite{xie2025foxm1}, with Xie and colleagues showing that disruption of FOXM1 condensates by AMPK agonists reduce breast-tumor growth and metastasis \cite{xie2025foxm1}. Stress-granule modulation has been proposed in cancer to mitigate chemotherapy resistance in tumors \cite{jia2024stress,wang2020targeting}, while targeting P-bodies has been considered in Parkinson's disease and selected cancers \cite{webber2020pathophysiology,hallacli2022parkinson,nsengimana2022processing}.
%
%
%
%

\begin{table*}[!ht]
\centering
\caption{Condensate-relevant therapies in registered clinical trials. Compounds are listed where the sponsor pipeline or peer-reviewed literature describes a mechanism of action that maps onto the colloidal-physics framework of Section~\ref{sec:Pathologies}. Phase, status, and sponsor reflect entries in ClinicalTrials.gov as of October, 2026.}
\label{tab:clinical}
\small
\begin{tabularx}{\textwidth}{@{}p{2.2cm}XXp{2.6cm}p{2.0cm}X@{}}
\toprule
\textbf{Compound} & \textbf{Mechanism (this review)} & \textbf{Indication} & \textbf{Sponsor} & \textbf{Highest stage} & \textbf{References} \\
\midrule
DPTX3186 &
Disruption of Wnt-pathway-activated transcriptional condensates; partition failure, mechanism~(c) &
Wnt-pathway-activated solid tumors (colorectal, gastric, lung, triple-negative breast) &
Dewpoint Therapeutics &
Phase~1, recruiting (first-in-human) &
NCT07312903 \\[0.5em]
Anle138b / emrusolmin (TEV-56286) &
$\alpha$-Synuclein oligomer modulator; aligned with the liquid-to-solid trajectory of mechanism~(b) &
Multiple system atrophy (MSA); an earlier Phase~1 trial in Parkinson's disease &
Teva Branded Pharmaceutical Products/ MODAG GmbH &
Phase~2, active &
NCT06568237, NCT04685265 \cite{wagner2013anle138b} \\
\bottomrule
\end{tabularx}
\end{table*}

Therapeutic targeting of the remaining three mechanism classes is at an earlier proof-of-concept or conceptual stage. For pathogen hijacking (mechanism (e)), Risso-Ballester and colleagues demonstrated condensate \emph{hardening} as an antiviral strategy: the small molecule A3E and the steroidal alkaloid cyclopamine disorganize and harden respiratory-syncytial-virus inclusion-body condensates, inhibiting viral RNA synthesis and producing antiviral activity in mouse models \cite{rissoballester2021condensate}. The broader pathogen-hijacking mechanism class has motivated antiviral programs in which host or viral phase-separation contacts are blocked directly \cite{mullard2019biomolecular,vendruscolo2022protein}, but no compound has yet advanced to registered human trials. Similarly, for saturation concentration dysregulation (mechanism (a)), no condensate-specific compound has been advanced to clinical trials. Islet amyloid polypeptide (IAPP) in type-2 diabetes provides a biophysical illustration in which therapeutic approaches that reduce IAPP burden or restore secretory balance address this concentration driver indirectly \cite{mukherjee2015type,halban2014beta}. For interfacial aggregation (mechanism (d)), the mechanism class itself was characterized only over the past four years, and candidate interventions targeting the dense-dilute boundary --- surfactant-like proteins that compete with amyloidogenic adsorbers, RNA modulation that alters the interfacial $\zeta$-potential, and charge tuning of the condensate scaffold --- remain at the conceptual stage \cite{visser2025controlling}.

Condensate-targeted therapeutics face several open challenges that the above strategies have yet to fully resolve. First, selective partitioning remains difficult to engineer: a small molecule whose $K$ is high for a target condensate may have comparable $K$ in other condensates sharing similar IDR chemistry \cite{klein2020partitioning,mitrea2022modulating}, though simulation-based prediction of small-molecule partitioning is beginning to address this design problem \cite{emelianova2025prediction}. Second, the causal relationship between aberrant condensate behavior and disease progression is debated for several of the pathologies discussed in Section~\ref{sec:Pathologies}. Targeting a condensate that is downstream of the primary pathological cascade may merely ameliorate a biomarker without addressing the underlying cause. Third, the clinical translation of condensate-targeting strategies is at an early stage: as of 2026, the registered-trial inventory remains small, with the compounds in Table~\ref{tab:clinical} representing the frontier of a largely preclinical field. These limitations argue for rigorous causal dissection of condensate involvement in each disease context alongside efforts to develop condensate-targeting therapeutic modalities.

\subsection{Condensates as drug carriers and engineered platforms}\label{sec:TherapCarriers}
Condensates themselves, due to their tunable biophysical properties, are actively explored as drug delivery vehicles (Fig.~\ref{fig:condensatetherapeutic}b) \cite{liu2023peptide,sun2022phaseseparating}. Compared with established carriers such as micelles, liposomes, and polyplexes, coacervate vehicles offer several physical advantages: high payload partitioning (typical $K \sim 10^2$--$10^3$) and tunable interfacial tension ($\gamma \sim 10^{-7}$--$10^{-4}$ N/m) paired with tunable internal viscoelasticity that together set release kinetics \cite{sing2017development,klein2020partitioning,jawerth2018salt, wang2025rheology,law2023bending}. An ultralow $\gamma$ makes coacervates mechanically fragile relative to other delivery vehicles, but this limitation can be mitigated by Pickering-like surface stabilization with proteins or amphiphilic polymers, the same strategy that protects biological condensates from coalescence in cells \cite{folkmann2021regulation}.
 
Partitioning-based encapsulation strategies enable drug payloads to be loaded into coacervate carriers by deliberately matching the cargo's chemistry to maximize $K$ and then released by environmental triggers that alter $\Delta G_{\rm part}$ (Fig.~\ref{fig:condensatetherapeutic}b) \cite{liu2023peptide,sun2022phaseseparating}. pH gradients between extracellular and endosomal compartments, redox switches between extracellular oxidizing and intracellular reducing milieus, enzymatic cleavage of the polyelectrolyte backbone, and thermoresponsive transitions are all potential release triggers \cite{liu2023peptide,sun2022phaseseparating}. Because $\Delta G_{\rm part}$ enters Eq.~\ref{eq:Kpart} exponentially, modest shifts in environmental chemistry can also produce order-of-magnitude swings in the retained payload, paralleling the same physical principle by which post-translational modifications regulate biological condensate composition. 

A more recent direction designs synthetic condensates that operate inside cells (Fig.\ref{fig:condensatetherapeutic}c). Light-activated multivalent scaffolds, engineered IDR fusion proteins, and minimalistic reconstitution platforms allow researchers to nucleate condensates of controlled composition and lifetime in cellular environments \cite{shin2017optodroplets,bracha2018mapping,dai2023engineering,kim2022recent}. In vitro transcription-translation systems built around coacervate scaffolds extend the same approach to cell-free manufacturing, with polypeptide--DNA condensates serving as programmable reaction compartments \cite{schoenmakers2023vitro}. Applications under development include targeted local protein production, intracellular delivery of therapeutic RNA (Fig.\ref{fig:condensatetherapeutic}c), and engineered immune cells in which the two-dimensional signaling clusters introduced in Section~\ref{sec:Types2D} are tuned for enhanced specificity \cite{sun2022phaseseparating,liu2023peptide,schoenmakers2023vitro,xu2024phase}. Chimeric antigen receptor T-cell (CAR-T) engineering is a current example: receptor sensitivity and downstream signaling thresholds can be set by the phase behavior of the engineered receptor's intracellular domain \cite{xu2024phase}.

Across both threads, the controlling physical parameters are the same: the four axes $\{\chi, K, \tau_R, \gamma\}$ jointly determine whether a condensate forms, what it concentrates, how its material state evolves, and how its interface behaves. Targeting a disease and engineering a carrier are operationally distinct activities that share these parameters as control variables, and both benefit from understanding and characterizing $\chi$, $K$, $\tau_R$, and $\gamma$ for a system of interest.

\section{Conclusion and future directions}\label{sec:Conclusion}
Liquid-liquid phase separation of intrinsically disordered biomolecules is now established as a generic organizing principle of the cell, supported by direct measurement of the relevant physicochemical quantities across a growing number of model systems. This review has surveyed three-dimensional and membrane-tethered condensates (Section~\ref{sec:Types}), organized their physical chemistry around the colloidal, interfacial, and transport phenomena that govern condensate behavior (Sections~\ref{sec:Colloidal}--\ref{sec:Transport}), and identified four quantitative axes $\{\chi, K, \tau_R, \gamma\}$ that may be used to physically distinguish between functional and pathological condensates. These axes may also be used to guide the therapeutic and engineering applications developed in Sections~\ref{sec:Pathologies} and~\ref{sec:Therapeutic}.

Viewing the field from this physical perspective has three concrete benefits. First, the same physical quantities (in this case $\{\chi, K, \tau_R, \gamma\}$) used for cytoplasmic three-dimensional droplets can be adapted to describe membrane-tethered two-dimensional clusters, enabling a cohesive understanding of condensates across dimensions and topologies. Second, these physical parameters that distinguish a functional condensate from a pathological one can also be used to guide the design of condensate-targeted therapies and condensate-based drug carriers, presenting a unified lens to view the condensate-pathology-therapy landscape. Finally, the framework links condensate biology to classical soft matter in a mutually beneficial manner: biology gains quantitative and theoretical tools, while soft matter benefits from an effectively infinite sequence-encoded design space of biopolymers that can drive phase separation, opening routes to novel synthetic programmable materials.

The colloidal, interfacial, and transport perspective presented here consolidates a substantial body of condensate research into a small set of physical relationships, and it should be carefully employed with an understanding of the limitations. First, the Flory-Huggins theory (Eq.~\ref{eq:fh}) treats the dense phase as a homogeneous mean field, yet condensates can have heterogeneous internal structure such as the FC/DFC/GC layering of the nucleolus \cite{feric2016coexisting,jaberi2023evolutionarily}. Further, pathologically aged condensates with heterogeneous irreversible cross-links can have multimodal relaxation times, transcending the single characteristic timescale expected for a simple Maxwell-like or Kelvin-Voigt description. Causal relationships between aberrant condensate behavior and disease also remain debated and unsettled for several of the pathologies discussed in Section~\ref{sec:Pathologies}. More fundamentally, much of the cellular condensate machinery is not at equilibrium -- ATP-driven dynamics, post-translational modifications, and active fluxes can transiently modify $\chi$, $K$, $\tau_R$, and $\gamma$ values that equilibrium forms only approximate \cite{berry2018physical}.

Several future research opportunities at the intersection of condensate biology and colloidal, interfacial, and transport physics can collectively advance the state of the art. The first is sequence-to-property prediction. Proteome-scale coarse-grained simulations can now generate conformational ensembles for the $\sim$28,000 intrinsically disordered regions of the human proteome and link their Flory scaling exponents to cellular function and subcellular localization \cite{tesei2024conformational}. Carrying that capability from single-chain conformation to collective condensate properties is an open problem, and advances in predicting partitioning coefficients illustrate both the progress and the gap: $\Delta G_{\rm part}$, which sets $K$ through Eq.~\ref{eq:Kpart}, is not yet sequence-predictable at the precision needed for therapeutic design, though sequence- and simulation-based predictors are advancing rapidly \cite{wessen2025sequence,ginell2025sequence,emelianova2025prediction}. For $\gamma$ and $\tau_R$, experiments have linked specific sequence changes to measured properties \cite{holland2023surface,alshareedah2024sequence}, but these relationships have not been generalized into predictive models. The second direction inverts this mapping. Forward prediction asks what a given sequence will do, whereas design requires specifying target values of $\chi$, $K$, $\tau_R$ and $\gamma$ and engineering sequences that realize them. Synthetic platforms already provide considerable control over condensate assembly, including light-activated scaffolds that set valence and concentration independently \cite{shin2017optodroplets,bracha2018mapping}, engineered condensates of programmed composition \cite{dai2023engineering}, and coacervates whose growth and dissolution are coupled to a designed reaction network \cite{patra2025programmable}, but their properties are read out after assembly rather than prescribed beforehand. Closing that loop would convert the four axes from descriptors into design variables, and such a capability can accelerate the therapeutic and materials applications of condensates. The third direction is to advance the understanding of two-dimensional condensates. Membrane-tethered condensates have a thermodynamic and interfacial description, with a surface saturation $\Gamma_{\rm sat}$ replacing $c_{\rm sat}$ and a line tension replacing $\gamma$ (Section~\ref{sec:Colloidal}). However, experimental characterization of two-dimensional condensates has been limited: line tension has not been determined for any protein condensate, and no membrane-tethered condensate has been characterized rheologically. The aberrant-aging and liquid-to-solid class of Section~\ref{sec:Pathologies} therefore has no established two-dimensional counterpart, and whether membrane-tethered condensates age at all remains unknown. The fourth direction is quantitative in-cell metrology of the condensate properties at submicron resolutions. Time-shared optical-tweezer microrheology can measure the viscoelastic response of intracellular compartments over a millipascal-to-kilopascal modulus range and five decades of frequency with reported measurements inside zebrafish progenitor and \textit{C.~elegans} intestinal cells \cite{catala2025measuring}, and high-pressure-freezing combined with cryo-focused-ion-beam milling and context-aware template matching now resolves nucleosome packing inside native chromatin condensates at near-nanometer detail \cite{zhou2025quantitative}. Building similar pipelines for measuring condensate properties inside cells would convert isolated results into proteome-wide datasets, underpinning every other direction discussed above. Together, these programs would deliver a quantitative biology of condensates that is physically grounded and would in turn accelerate the engineering and therapeutic goals of the field.

\section*{Acknowledgments}
We acknowledge the use of \url{https://www.biorender.com} in creating schematics. During the preparation of this work the authors used Gemini (Google LLC) and Claude (Anthropic) to check grammar and improve readability. After using these tools, the authors reviewed and edited the content as needed and take full responsibility for the content of the published article.

\bibliographystyle{vancouver}
\bibliography{Reference}

\end{document}